%% file: nmc_arxiv_verfin.tex
\documentclass[reprint,twocolumn,floatfix,aps,showpacs,footinbib,amsmath,amssymb,amsfonts,superscriptaddress]{revtex4-2}
\usepackage{natbib}
\usepackage{url}
\usepackage{orcidlink}
\usepackage{graphicx}
\usepackage{color}
\usepackage{lipsum, babel}
\usepackage{hyperref}
\usepackage{mathtools}
\usepackage{mathrsfs}
\usepackage{calrsfs}
\usepackage{comment}
\usepackage[utf8]{inputenc}
\usepackage{multirow} 
\usepackage{subfigure}
\usepackage{varioref}
\usepackage[active]{srcltx}
\hypersetup{
	colorlinks=true,       
	linkcolor=blue,          
	citecolor=blue,        
}

\usepackage[section]{placeins}

\usepackage{fancybox}
\labelformat{figure}{Fig.~#1}

\newcommand*{\Scale}[2][4]{\scalebox{#1}{$#2$}}%

\begin{document}

\title{A test of Einstein's equivalence principle in future VLBI observations}

 \author{Joseph P Johnson\orcidlink{0000-0003-4618-2092}}
 \email{josephpj@iisermohali.ac.in}
 \affiliation{Department of Physics, Indian Institute of Technology Bombay, Mumbai 400076, India}
 \affiliation{IISER Mohali, Knowledge City, Sector 81, SAS Nagar, Manauli PO 140306, India}
 \author{Susmita Jana\orcidlink{0000-0001-7681-7533}}
 \email{susmitajana@iitb.ac.in}
 \affiliation{Department of Physics, Indian Institute of Technology Bombay, Mumbai 400076, India}
 \author{S. Shankaranarayanan\orcidlink{0000-0002-7666-4116}}%
 \email{shanki@iitb.ac.in}
 \affiliation{Department of Physics, Indian Institute of Technology Bombay, Mumbai 400076, India}

 \begin{abstract}
 We show that very-long-baseline-interferometry (VLBI) observations of supermassive black holes will allow us to test the fundamental principles of General Relativity (GR). GR is based on the universality of gravity and Einstein's equivalence principle (EEP). However, EEP is not a basic principle of physics but an empirical fact. Non-minimal coupling (NMC) of electromagnetic fields violates EEP, and their effects manifest in the strong-gravity regime. Hence, VLBI observations of black holes provide an opportunity to test NMC in the strong-gravity regime. To the leading order in the spin parameter, we explicitly show that the NMC of the electromagnetic field introduces observable modifications to the black hole image. {In addition, we find that the size of the photon rings varies by $\sim 3 r_H$, which corresponds to $\sim 30 \mu as$ for Sagittarius $A^*$ and $\sim 23 \mu as$ for M87.
 VLBI telescopes are expected to attain a resolution of $\sim 5 \mu as$ in the near future. However, direct detection of photon ring will require the resolution of $\sim 1 \mu as$ for M87, which can potentially be probed by the space-based Event Horizon Explorer}. 
 \end{abstract}
\maketitle

\noindent \emph{{Introduction:}} Einstein's general relativity (GR) has been tested at solar system scales and in the weak field limit~\cite{2014-Will-LivingRev.Rel.,2015-Berti-Class.Quant.Grav.,2018-Will-Book}. Until recently, strong-gravity regime observations were difficult.  
The detection of gravitational waves by the LIGO-VIRGO-KAGRA collaboration~\cite{2019-Abbott-Phys.Rev.D,2021-Abbott-Phys.Rev.D,2021-Abbott-} and the imaging of the black hole shadow by the Event Horizon Telescope (EHT) using Very-Long-Baseline
Interferometry (VLBI)~\cite{2019-Akiyama-Astrophys.J.Lett.,2022-Akiyama-Astrophys.J.Lett.,2022-Akiyama-Astrophys.J.Lett.a} offer fresh possibilities to search for deviations from GR in the strong-gravity regime. 

GR is based on the universality of gravity and {Einstein's equivalence principle (EEP)}~\cite{2018-Will-Book,2017-Misner-Gravitation}. {EEP} specifies the interaction between gravity and all other matter fields~\cite{1960-Schiff-AmJP,2019-Dicke-GRG,1973-Thorne.etal-PRD}.
EEP is divided into three sub-principles: the weak EP (the universality of free fall), local Lorentz invariance, and local position invariance. EEP is satisfied if all these three sub-principles are satisfied~\cite{2018-Will-Book}. 

However, EEP is not a basic principle of physics but an empirical fact~\cite{2001-Damour-Talk}. Hence, one way to assess the validity of GR is to test EEP. 
In the weak field limit, the weak EP has been tested in several parts in $10^{16}$ using the torsion balance, lunar-laser ranging, and space missions like MicroSCOPE~\cite{1991-Damour.Schaefer-PRL,1996-Williams.etal-PRD,1999-Baessler.etal-PRL,2001-Barkovich.etal-PLB,2006-Kesden.Kamionkowski-PRD,2007-Schlamminger.etal-PRL,2010-Keselman.etal-PRD,2012-Wagner.etal-CQG,2012-STEP-CQG,2013-Overduin.eta-CQG,2014-Icecube-PRD,2016-Wang.etal-PRL,2017-Microscope-PRL}. Experiments using laser-cooled trapped atoms to look for variations in the relative frequencies of different types of atoms as the Earth rotates around the mean rest frame of the universe have placed very strict constraints on local Lorentz invariance~\cite{2005-Mattingly-LRR,2013-Liberati-CQG,2014-Will-LivingRev.Rel., 2015-Kislat-Phys.Rev.D}. Additionally, gravitational redshift experiments and tests of variations in fundamental constants have been used to examine local position invariance~\cite{2011-Uzan-LRR}.

While the tests of EP have improved over the years, contamination by complex physics is one of the primary challenges of testing GR in the strong-gravity regime. This work aims to develop an observational test of GR by investigating the \emph{potential violation} of EEP 
in future VLBI observations.  
Although the black hole image depends on the details of the accretion flow, two signatures of the image are unique~\cite{2008-Psaltis-LRR}:
First, regardless of the black hole's spin, the horizon casts a shadow on the image of the source at around $\sqrt{27} GM/c^2$~\cite{2000-Falcke.etal-ApJL}. Second, due to the high velocity of the accreting plasma and the effects of gravitational lensing, the image brightness of the accretion flow is highly non-uniform~\cite{2019-Gralla-Phys.Rev.D}. 
We show that non-minimal coupling (NMC) of the electromagnetic field distinctly modifies these two signatures
~\cite{1971-Prasanna-PhysicsLettersA,1974-Deser-Phys.Rev.D,1976-Horndeski-JMP,1980-Drummond-Phys.Rev.D,2002-Feynman-Book}.

EEP excludes products between curvature and matter fields (aka NMC) in action~\cite{2022-Shanki.Joseph-GRG}. One way EEP violation occurs is due to NMC~\cite{1971-Prasanna-PhysicsLettersA,1974-Deser-Phys.Rev.D,1976-Horndeski-JMP}. Interestingly, NMC is required for consistency of the standard model in curved space-time~\cite{1980-Drummond-Phys.Rev.D,2002-Feynman-Book}. NMC is unique if we demand at most second derivatives of $g_{\mu\nu}$, and the electromagnetic potential ($A_{\mu}$) appears in the field equations~\cite{1976-Horndeski-JMP}. The effects due to NMC manifest in the strong-gravity regime, such as in the cores of neutron stars or close to the black hole horizon. The curvature effects cannot be ruled out despite the extreme precision of current experiments on the electromagnetic field~\cite{1973-Hellings.Nordtvedt-PRD,2015-Kislat-Phys.Rev.D,2022-Kostelecky.Mewes-PRD}.

 VLBI observations of black holes allow for testing NMC as their effects are substantial near black hole horizons~\cite{2023-Vagnozzi-Class.Quant.Grav.}. EHT image shows glowing gas circling the black hole at high speed and a black hole shadow caused by lensing and photon capture~\cite{2019-Akiyama-Astrophys.J.Lett.,2022-Akiyama-Astrophys.J.Lett.a}. Future VLBI telescopes, including the space-based projects~\cite{2022-Kurczynski-} with the help of advanced data analysis techniques, and imaging techniques such as super-resolution are expected to detect and observe the unstable photon circular orbit as a photon ring ~\cite{2020-Broderick-apj,2020-Johnson-Sci.Adv.,2021-Hadar-Phys.Rev.D, 2023-Roelofs-Galaxies}. Furthermore, as we show explicitly, NMC will lead to modified dispersion relations leading to \emph{different propagation} for the two polarization modes. We also analyze the difference in the resulting black hole shadow for the two modes compared to GR for Schwarzschild and slowly rotating Kerr.

First, we analyze the photon trajectories in the Schwarzschild black hole and 
observe photon flux distribution for each mode that determines the properties of the black hole image. We show that, for one polarization mode, the horizon casts a shadow of radius \emph{greater than} $\sqrt{27} GM/c^2$ on the source image. For the other polarization mode, it is \emph{smaller than} $\sqrt{27} GM/c^2$.
Later, in the case of the Kerr black hole, we show that the size of the observed black hole shadow is largely independent of the spin of the black hole. The effect of NMC on the shadow size follows a similar trend as in the case of Schwarzschild. [By shadow, we mean the interior region of the observed photon ring.] Finally, we discuss the possibility of constraining the NMC constant using future VLBI observations.

[Variables, like $({r}, t, {s}, {a})$, \emph{used below are dimensionless}. They are related to dimensionful variables $(\tilde{r}, \tilde{t}, \tilde{s}, \tilde{a})$
by a factor $1/{r}_H$ where ${r}_H = 2 GM/c^2$ is the Schwarzschild radius. The rotational parameter ($a$) is twice the dimensionless rotational parameter used in the literature. {\it Over-dot} denotes derivative w.r.t. dimensionless affine parameter $s$ and {\it prime} denotes derivative w.r.t. $r$.]

\noindent \emph{Implications of NMC for Schwarzschild:} As mentioned above, NMC is unique if we demand at most second derivatives of $g_{\mu\nu}$ and $A_{\mu}$ appear in the field equations~\cite{1976-Horndeski-JMP}. For the Ricci-flat space-times, NMC electromagnetic (test) field action is~\cite{1971-Prasanna-PhysicsLettersA,2003-Prasanna-Class.Quant.Grav.}
\begin{equation}
\mathrm{S}= \int d^4 x \sqrt{-g}\left[
   - F_{\mu \nu} F^{\mu \nu} + 2  {\tilde{\lambda}} R^{\mu \nu \alpha \beta} F_{\mu \nu} F_{\alpha \beta} \right]/4 \, ,
\end{equation}
where $\tilde{\lambda}$ is the dimensionful NMC constant. The resulting equation of motion is:
\begin{equation}
\nabla_\nu F^{\mu \nu}= 2 \tilde{\lambda}\left[R^{\mu \nu \rho \sigma}\left(\nabla_\nu F_{\rho \sigma}\right)+\left(\nabla_\rho R_\sigma^\mu\right) F^{\rho \sigma}\right] \, .
\end{equation}
In the local inertial (tetrad) frame basis, the above equation leads to the following dispersion relation~\cite{2003-Prasanna-Class.Quant.Grav.}:
\begin{equation}
\label{eq:dispersion-general-metric}
\left[p^{(\mu)} p_{(\mu)} \delta^{(j)}_{(k)} + 4\tilde{\lambda} \left[\frac{p^{(j)}}{p_{(0)}} \epsilon^{(0)}_{(k)}  
+ \epsilon^{(j)}_{(k)} \right]  
\right] F^{(0)(k)} = 0,
\end{equation}
where $p_{(\mu)}$ is the photon momenta in the tetrad basis and 
\begin{equation}
\epsilon^{(\alpha)}_{(\beta)} \equiv R^{(\alpha)(\mu)(\nu)}\, _{(\beta)} \, p_{(\mu)}p_{(\nu)} \, .
\end{equation}
Note that the above expression is valid for {any} Ricci-flat space-time. However, the tetrad basis depends on the background geometry, leading to modified dispersion relations for different space-times. 

To discern the effects of NMC on the photon orbit and the flux, we first consider the Schwarzschild background:
\begin{equation}
ds^2 = 
- f(r) \, dt^2 + dr^2/f(r) + r^{2} d\Omega^2 \, , 
\end{equation}
where, $f(r) := 1 - 1/r$ and $d\Omega^2$ is the metric on unit $\mathbb{S}^2$.  In these coordinates, the conserved quantities --- energy ($E$) and angular momentum ($L$) --- are:
\begin{equation}
    \label{def:E-L-Sch}
E = f(r) \, \dot{t} 
\, , \quad
L = r^2 \, \dot{\phi} 
\, .
\end{equation}
Substituting the above metric in Eq.~\eqref{eq:dispersion-general-metric} and using the tetrad basis defined (in Eq.(31.4a)) in Ref.~\cite{2017-Misner-Gravitation}, the modified dispersion relation  for the two polarization modes of the photon (denoted by $'+'$ and $'-'$) is:
\begin{equation}
\label{eq:sch-disp}
\!\! p^2 
= \mathcal{C}_{\pm}\, p_{(\phi)}^2,~ \mathcal{C}_+ = 
\dfrac{6 {\lambda}}{2 {\lambda}+{r}^3},~\mathcal{C}_- = \dfrac{6 {\lambda}}{4 {\lambda} - {r}^3},
\end{equation}
where $p^2 = p^{\mu} p_{\mu}$, and ${\lambda} = \tilde{\lambda}/ r_H^2$ is dimensionless. In the {equatorial plane}, Eq.~\eqref{eq:sch-disp} reduces to:
\begin{equation}
\label{eq:sch-rdot}
    \dot{r}^2= {E}^{2}\, \left[1 - {{b}^2} \left(1 - \mathcal{C}_{\pm} \right) f(r)/r^2  \right],
\end{equation}
where ${b} = {L}/{E}$ denotes the impact parameter. 
The photon circular orbit satisfy two independent conditions~\cite{1983-Chandrasekhar-MathematicalTheoryBlack,1972-Bardeen-Astrophys.J.,2013-Bambi-Phys.Rev.D}: $\dot{{r}}=0$ and $\ddot{{r}}=0$.
It is not easy to obtain an exact analytical solution for these equations; hence, we numerically obtain the radius (${r}_c$) of the circular orbits and the corresponding impact parameter ${b}_c$. (Using the perturbative approach, one can obtain $r_c$ and $b_c$. These are consistent with numerical results. For details, see the supplementary material~\cite{sup1-}.)

\ref{fig:sch_rb} contains the plot of $r$ (solid line) and $b$ (dotted line) {corresponding to the circular orbit} for the $+, -$ modes as functions of $\lambda$.  {For a Schwarzchild black hole, $b_c$ denotes the position of the photon ring as seen by a distant observer}. As mentioned, we consider the interior of the photon ring as the shadow of the black hole. 
 However, in general scenarios, depending on the characteristics of the accretion process that acts as the light source around the black hole, the observed shadow consists of a region larger than the photon ring. The black hole shadow with NMC for Novikov-Thorne accretion model~\cite{1973-Novikov-} (with low numerical accuracy) is discussed in the supplementary material \cite{sup1-}. Figs 2 and 3 in \cite{sup1-} contain black hole image for the two accretion flows.
\begin{figure}[!htb]
\includegraphics[scale=0.30]{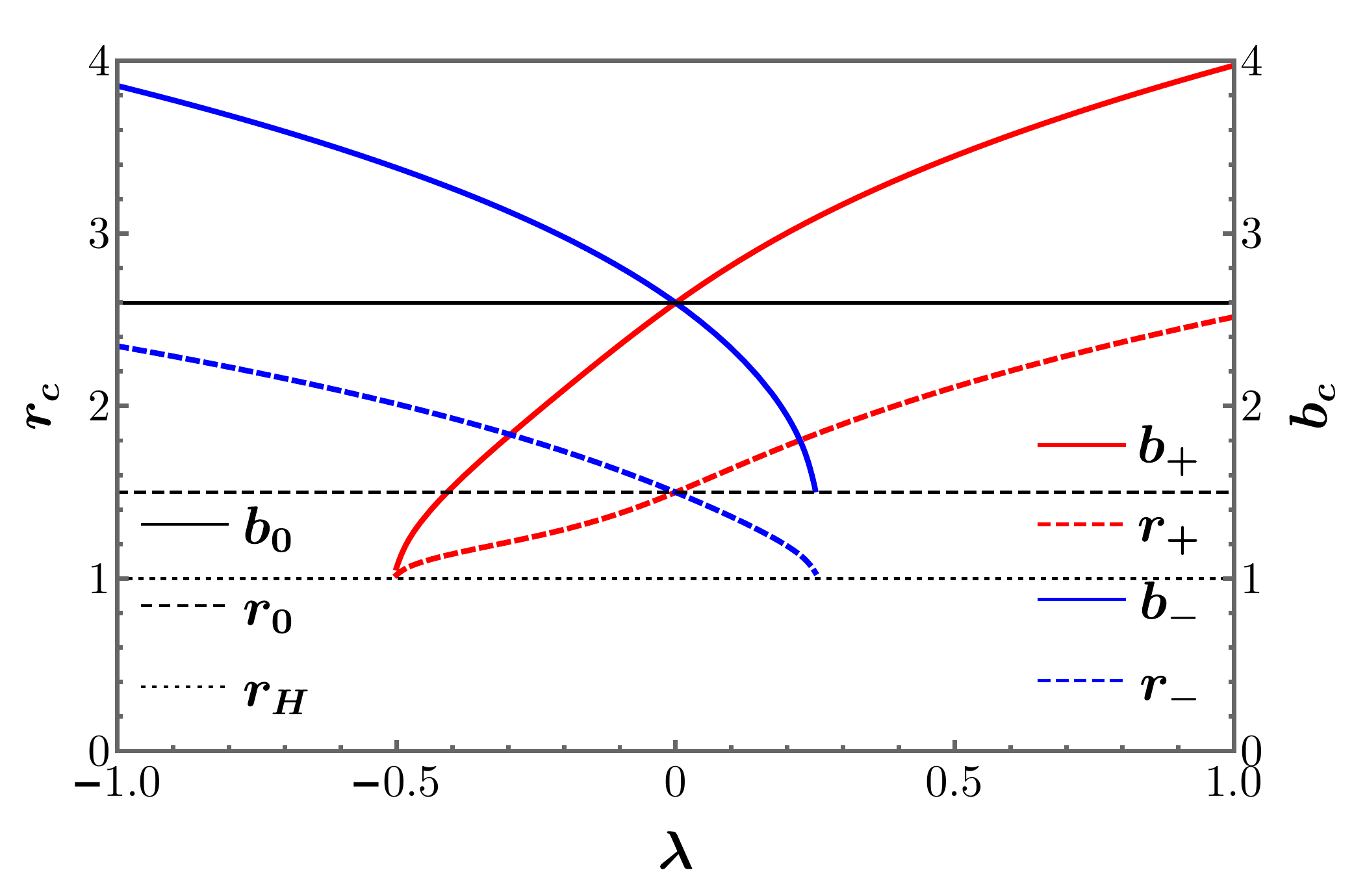}
\caption{Position of the photon ring $r_{\pm}$ and corresponding impact parameter $b_{\pm}$ as functions of the coupling constant $\lambda$ along with the values corresponding to the minimally coupled (MC) mode ($r_0, b_0$) and the horizon radius $r_H$.}
\label{fig:sch_rb}
\end{figure}

From the above figure, we infer the following: 
First, for both ($'+'$ and $'-'$) modes, $r_c$ and 
$b_c$ intersect at $\lambda = 0$. 
Second, there is no observable photon ring ($r_c>r_H$) for $'+'$ mode in the case of $\lambda< -0.5$ and $'-'$ mode in the case of $\lambda > 0.25$.
%
Third, for $'+'$ mode, when ${\lambda}>0$ radius of the photon sphere and the corresponding impact parameter are larger than the minimally coupled case, resulting in a larger observed shadow. When ${\lambda} < 0$, $'+'$ mode forms a smaller shadow. This trend is reversed in the case of $'-'$ mode, where ${\lambda}>0$ leads to the formation of a smaller shadow, and ${\lambda} < 0$ results in a larger shadow. 
Lastly, the perturbative approach to obtain $r_c$ and $b_c$ matches with numerical results for $|\lambda| < 0.2$ (See supplementary material \cite{sup1-}). Later, we use this approach for Kerr to obtain $r_c$ and $b_c$. As we show below, the above results for Schwarzschild continue to hold for Kerr. 
%

{The probability of photons from the unstable circular orbit reaching the distant observer, hence forming a bright ring in the observer's image plane with its position given by $b_c$~\cite{1972-Bardeen-Astrophys.J.,1983-Chandrasekhar-MathematicalTheoryBlack}.} We now obtain the flux for the NMC photons by considering radially free-falling optically thin light-emitting accretion flow~\cite{2000-Falcke.etal-ApJL}.
For such a system, photon flux as observed at a point $(x,y)$ in the observer's image plane is~\cite{2013-Bambi-Phys.Rev.D}:
\begin{equation}
F(x,y) \propto \int dr~g^3 { p_t}/[|p_r| r^2], 
\end{equation}
where $g \equiv {p_{\alpha} \, u^{\alpha}_{\rm obs}}/({p_{\beta} u^{\beta}_e})$ is the red-shift factor. $u_{\rm obs}$ and $u_e$ are the four velocities of distant observer and accreting gas,
respectively, and are given by:
\begin{equation}
u_{\rm obs}=(1,0,0,0),~~
u_e = \left(1/f(r), 
-{1}/{\sqrt{{r}}},0,0\right) \, .
\end{equation}
The impact parameter of the photon trajectory is related to the image plane coordinates by ${b}^2 = x^2 + y^2$. 
\begin{figure}[!htb]
\centering
\includegraphics[scale=0.18]{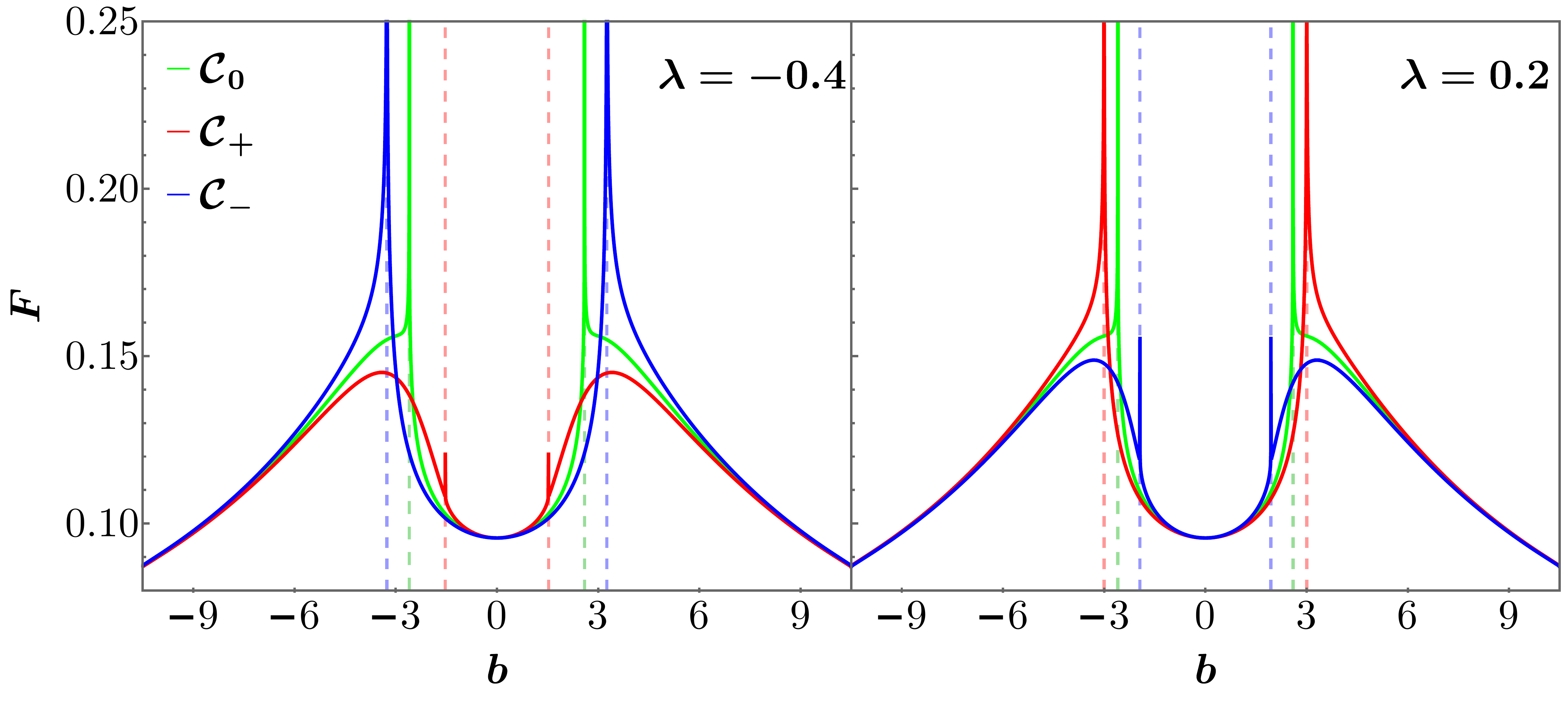}
\caption{Observed flux as a function of impact parameter $b$ for $'\pm' ({\cal C}_{\pm})$, and minimally coupled $({\cal C}_0)$ modes. The dashed lines are the position of the photon ring (cf. \ref{fig:sch_rb}).}
\label{fig:flux_b}
\end{figure}

In \ref{fig:flux_b}, we plot the observed flux as a function of $b$ for two specific values of $\lambda$.
 %
From the figure, we infer the following: First, the spikes in the observed flux (scaled by a constant) correspond to the observed photon ring for the two NMC modes and minimally coupled photons.
Second, in the case of minimally coupled photons, it is known that the horizon casts a \emph{shadow} on the image of the source at $\sqrt{27}~GM/c^2 = \sqrt{27/4} \, r_H$~\cite{2000-Falcke.etal-ApJL}. However, for ${\lambda}>0$ and $'+'$ mode, the shadow radius is greater than $\sqrt{27/4} \, r_H$ and for $'-'$ mode, the shadow radius is smaller. This trend is reversed for ${\lambda}<0$. 
Third, we have not imposed the circular orbit conditions to obtain the flux distribution. These results, including the observed position of the photon ring, are consistent with \ref{fig:sch_rb}.
Lastly, we also see that the brightness and the position of the lensing ring are affected by the non-minimal coupling~\cite{2019-Gralla-Phys.Rev.D}. 
For ${\lambda}<0$ and $'+'$ mode (and ${\lambda}>0$ and $'-'$ mode), the lensing ring is more prominent as compared to other cases. Hence, the NMC leads to two distinct observable effects in Schwarzschild --- shadow radius changes and lensing ring characteristics are modified. 

An attentive reader might ask: How do the above results translate for Kerr space-time, describing the super-massive black holes at the galactic centers? As mentioned earlier, regardless of the black hole's spin, the horizon casts a shadow on the image of the source at around $\sqrt{27/4}~r_H$~\cite{2000-Falcke.etal-ApJL}. Noting that the black hole spin is not tightly constrained~\cite{2020-Fragione-Astrophys.J.Lett.,2022-Fragione-Astrophys.J.Lett.}, we show that the shadow radius of NMC photons in a slowly rotating black hole shows the same trend as Schwarzschild space-time. 

\noindent \emph{Implications of NMC for Kerr:} In dimensionless rational polynomial coordinates $(t,r,\zeta [= \cos \theta],\phi)$, the Kerr space-time is~\cite{2007-Visser}:
{\small
\begin{eqnarray}
\mathrm{d} {s}^2 &=& -\dfrac{\Delta}{\rho^2}\left[ \mathrm{d}{t} -a (1-\zeta^2)\mathrm{d}{\phi} \right]^2 + \frac{\rho^2}{\Delta} \mathrm{~d} {r}^2  + \frac{\rho^2}{1-\zeta^2} \mathrm{d} \zeta^2 \nonumber \\
&+& \dfrac{(1-\zeta^2)}{\rho^2}\left[ (r^2 + a^2) \mathrm{d}{\phi} - a \mathrm{d}{t} \right]^2  \, ,
\end{eqnarray}
}
where ${a}$ is dimensionless spin parameter, $\Delta={a}^2 + {r}^2 - {r}$, and $\rho^2 = {r}^2 + {a}^2 \zeta^2$. 
Substituting the above metric in Eq.~\eqref{eq:dispersion-general-metric} and using the tetrad basis defined in Eq. (2.9) in Ref.~\cite{1998-Saijo-Phys.Rev.D}, the modified dispersion relation for the two modes of the photon, for arbitrary $a$, is:
\begin{equation}
    {p^2} = \dfrac{6 (p_{(\zeta)}^2 + p_{(\phi)}^2) {\lambda}\rho^3}{3 {\lambda}\rho^3 \mp \sqrt{\rho^{12} + 9 {\lambda}^2 \rho^6 + 2 {\lambda} r \Gamma [ \rho^2  - 4 {\lambda} {r} \Gamma ]}} \, ,
\end{equation}
where $\Gamma = (3 \rho^2 - 4 {r}^2)$ and the relation between momenta in tetrad and coordinate bases is:
\begin{align}
    & p^2 = p_{{r}}^2 \dfrac{\Delta}{\rho^2} - \dfrac{({a}^2{E}-{a}{L}+{E}{r}^2)^2}{\Delta \rho^2} + p_{(\phi)}^2+p_{(\zeta)}^2, \nonumber \\  
    \label{eq:Disper-Kerr}
    & p_{(\phi)}^2+p_{(\zeta)}^2 =  p_{\zeta}^2 \dfrac{1-\zeta^2}{\rho^2} +  \dfrac{({L}-{a}{E}(1-\zeta^2))^2}{(1-\zeta^2)\rho^2} 
    .
\end{align}
%
Unlike the minimally coupled case, the above dispersion relation lacks separability in $r$ and $\zeta$. However, this does not affect the analysis of the photon trajectory in the equatorial plane.
For the $'\pm'$ photon modes, photon trajectory in the equatorial plane is determined by
\begin{equation}
\!\!\!\! \frac{\dot{{r}}_{\pm}^2}{E^2} =  1 + \dfrac{{a}^2 - {b}^2}{{r}^2} + \dfrac{({a} - {b})^2}{{r}^2}\left[\frac{1}{r} + \mathcal{C}_{\pm} \left[ \dfrac{{a}^2}{{r^2}}  - f(r) \right]  \right].
\end{equation}
Like in the Schwarzschild, the above equations can not be solved analytically for the photon circular orbit. However, we can perturbatively solve the same by expanding the variables \emph{to any order} in $\lambda$ and $a$. As shown in 
In the ${a} \rightarrow 0$ limit, the perturbative solutions are consistent with the above results for Schwarzschild (see supplementary material~\cite{sup1-}). In ${\lambda} \rightarrow 0$ limit, the perturbative solutions reduce to Kerr results for minimal coupling{~\cite{1983-Chandrasekhar-MathematicalTheoryBlack}.}

\noindent \emph{Black hole shadow by a slowly-rotating Kerr:}
One must look at the photon trajectory around the black hole to analyze the shadow properties. Since the modified dispersion relation for Kerr \eqref{eq:Disper-Kerr} is not variable separable, obtaining analytical solutions for the photon trajectory is difficult. Interestingly, for a slowly rotating black hole at linear order in the spin parameter ($a$), the dispersion relation \eqref{eq:Disper-Kerr} can be separated and are given by:
\begin{multline}
  p_{\theta}^2 + \dfrac{{b}^2 E^2 }{\sin^2\theta} =  \dfrac{E^2\mathcal{C}_{\mp}}{f(r)}\left[ 2 a b \left[1-\dfrac{1}{r\mathcal{C}_{\mp}}\right] - \dfrac{r^2}{\mathcal{C}_{\pm}}\right] \\ 
  +p_{{r}}^2 \dfrac{\mathcal{C}_{\mp}}{\mathcal{C}_{\pm}} {r}^2 f(r) = k_{\pm} {E}^2,
\end{multline}
%
%
where ${\cal C}_{\pm}$ are defined in Eq.~\eqref{eq:sch-disp} and $k_{\pm}$ are the separation constants. 
Corresponding radial motion of the non-minimally coupled photon ($'\pm'$ modes) is:
{
\begin{equation}
\!\! \dot{{r}}_{\rm \pm}^2 =E^2\left[ 1+ \dfrac{\mathcal{C}_{\pm}}{r^2\mathcal{C}_{\mp}}\left[k_{\pm}f(r)  - 2 a b \left( \mathcal{C}_{\mp} - \dfrac{1}{r} \right)\right]\right] .
\end{equation}}
As in Schwarzschild space-time, we can obtain the circular orbit of the photon by imposing the conditions  
$\dot{r}=0$ and $\ddot{r}=0$. For the two modes, we get the following: 
\begin{eqnarray}
\nonumber
\Scale[1.0]{b_{\pm} = \dfrac{r^4 f(r)  [\mathcal{C}_{\mp} \mathcal{C}_{\pm}' - \mathcal{C}_{\pm} \mathcal{C}_{\mp}'] - r^2 [2rf(r) - 1] \mathcal{C}_{\pm} \mathcal{C}_{\mp}}{2 a {\mathcal{C}_{\pm}}^2(\mathcal{C}_{\mp} -r^2 f(r)  \mathcal{{C}_{\mp}' }-1)} \Bigr\rvert_{r=r_c}} \!\!\! && \\  
\label{eq:mpb}
\Scale[1.0]{k_{\pm} = \dfrac{r^2 [r\mathcal{C}_{\pm}\mathcal{C}_{\mp}'+r^2 f(r) \mathcal{C}_{\mp}\mathcal{C}_{\pm}'+\mathcal{C}_{\mp}\mathcal{C}_{\pm}[3- 2 r \mathcal{C}_{\mp}]]}{{\mathcal{C}_{\pm}}^2(\mathcal{C}_{\mp} -r^2 f(r)  \mathcal{{C}_{\mp}' }-1)} \Bigr\rvert_{r=r_c}} \!\!\!\!\!\! . &&
\end{eqnarray}
The above values of ${b}$ and $k$ 
(in terms of the radius $r_c$ of the circular orbits) determine the shape of the black hole shadow on the image plane of a distant observer. Before we proceed with the black hole shadow, we want to mention the following points about the validity of the slowly-rotating approximation for the supermassive black holes: First, the spin of the supermassive black holes observed by EHT is not tightly constrained~\cite{2020-Fragione-Astrophys.J.Lett.,2022-Fragione-Astrophys.J.Lett.}. Second, for minimal coupling, the horizon casts a \emph{shadow} on the image of the source at around $\sqrt{27/4}~r_H$~\cite{2000-Falcke.etal-ApJL}, regardless of the black hole's spin. 


%

Image of the photon ring seen by a distant observer can be described by the celestial coordinates $(x,y)$~\cite{1972-Bardeen-Astrophys.J.,1983-Chandrasekhar-MathematicalTheoryBlack}. These celestial coordinates are defined using the momentum components of photon measured by a distant locally non-rotating observer (LNRO)~\cite{1972-Bardeen-Astrophys.J.,1983-Chandrasekhar-MathematicalTheoryBlack}:
\begin{equation}
   x = -{\left[r {p^{(\phi)}}/{p^{(t)}}\right]}_{r \rightarrow \infty}, \quad y =  {\left[r {p^{(\theta)}}/{p^{(t)}}\right]}_{r \rightarrow \infty}.
\end{equation}
Using the relation between the non-zero components of the inverse tetrad in Boyer-Linquist coordinate  and LNRO frames (cf. Eq.(3.2) in Ref.~\cite{1972-Bardeen-Astrophys.J.}), the celestial coordinates in the distant LNRO's image plane are:
\begin{equation}
\label{eq:celcoord}
    x = - {b}/{\sin{\theta_0}}, \quad y = \pm\sqrt{k - b^2/\sin^2\theta_0} \, ,
\end{equation}
\begin{figure}[!htb]
    \centering
    \includegraphics[scale=0.45]{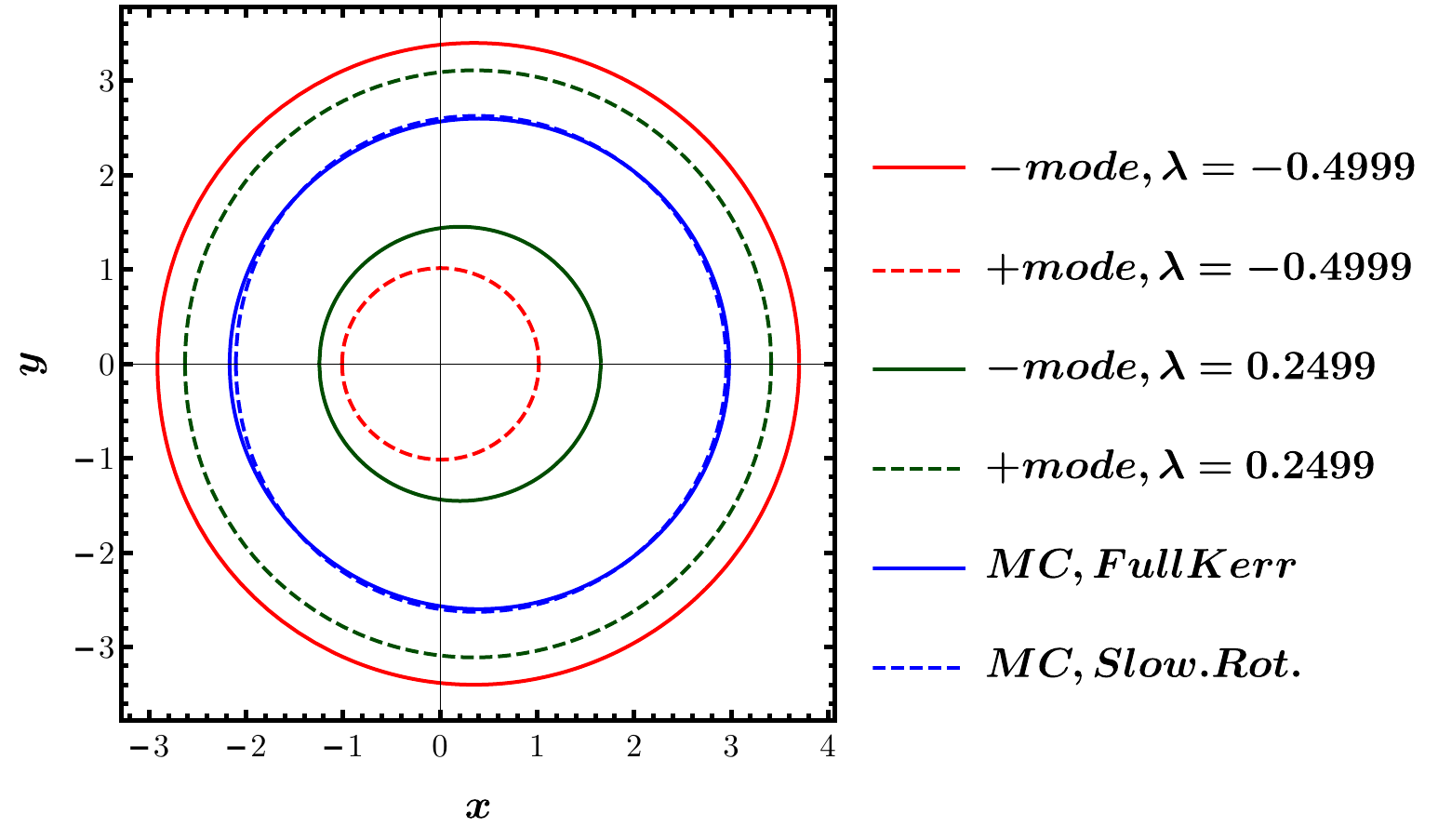}
    \caption{Outline of the shadow of a slowly-rotating black hole (${a}=0.2$) observed by a distant LNRO for $'\pm'$ modes for $\lambda = -0.4999, 0.2499,$ and $0$ (MC).}
    \label{fig:slow_rot_shadow}
\end{figure}
where $\theta_0$ is the inclination angle of the observer w.r.t. the symmetry axis of the black hole. Using Eq.~\eqref{eq:mpb}, one can evaluate the celestial coordinates $(x,y)$ corresponding to photon circular orbit with radius ${r}_c$. Collection of points $(x({r}_c), y({r}_c))$ for all possible values of $r_c$ will form the outline of the black hole shadow seen by a distant observer. The values of all possible ${r}_c$ are given by the conditions ${\rm Im}(y({r}_c)) = 0$ and ${r}_c \geq {r}_h$, where ${r}_h$ is the radius of the outer horizon of the black hole. 

\ref{fig:slow_rot_shadow} contains the outline of the shadow of a slowly rotating black hole observed by a distant LNRO for different photon modes for different $\lambda$ values. This is the key result of this work, regarding which we want to mention the following points: 
First, w.r.t ${\lambda}$, we see a similar trend as in the case of the Schwarzschild space-time. Observable photon circular orbits (${r}_c > {r}_h$) do not form if ${\lambda} \leq -0.5$ for $'+'$ mode and ${\lambda} \geq 0.25$ for $'-'$ mode. 
Second, for ${\lambda}>0$ the shadow radius is greater than $\sqrt{27/4}~r_H$ for the $'+'$ mode. The $'+'$ mode forms a smaller shadow when ${\lambda} < 0$. This trend is reversed in the case of $'-'$ mode; for ${\lambda}>0$ leads to the formation of a smaller shadow, and ${\lambda} < 0$ results in the shadow radius greater than $\sqrt{27/4}~r_H$.
Third, repeating the analysis for ${a} = 0.1$ and ${a} = 0.15$  shows that the spin ${a}$ does not have a significant effect on the shadow radius, even though there is a visible effect on the shadow shape (see Fig. 4 in supplementary material~\cite{sup1-}). This is similar to the results of the minimally coupled case for Kerr space-time~\cite{2000-Falcke.etal-ApJL}. Hence we expect that the predicted deviation in the shadow radius due to the non-minimal coupling will be similar for higher-spin black holes. However, this needs to be verified by numerics.

Fourth, {the current resolution of EHT} --- one of the popular VLBI telescopes --- has a resolution of $24 \mu a s$~\cite{2019-Akiyama-Astrophys.J.Lett.}. This translates to $2.4 r_H$ for Sagittarius $A^*$ and $3.2 r_H$ for M87,   in the image plane of the observer~\cite{2022-Akiyama-Astrophys.J.Lett.}. This is not good enough for the direct detection of the photon ring in the case of an equatorial accretion disk with the inner radius of $r_{ISCO}=3 r_H$. However, as discussed below, the resolution is expected to improve in future missions. 

Lastly, the difference between the shadow widths for the non-minimally coupled modes and minimally coupled case, as well as the difference between the $'+'$ and $'-'$ modes for the two nearly limiting values of ${\lambda}$, are: \\[1pt]
\noindent 1. Deviation of shadow size for $\lambda = -0.4999$
\begin{equation*}
   \Delta x_{-,0}=1.55, \quad  \Delta x_{+,0} = -3.03, \quad \Delta x_{+,-} = -4.58 .
\end{equation*}
2. Deviation of shadow size for $\lambda = 0.2499$
\begin{equation*}
    \Delta x_{-,0}=-2.16, \quad \Delta x_{+,0} = 0.98, \quad \Delta x_{+,-} = 3.14 .
\end{equation*}
Thus, we see that even though the current resolution of EHT is insufficient to directly detect the photon ring, it is expected that the future VLBI missions will be able to constrain the non-minimal coupling. Proposed projects include the Event Horizon Explorer, which aims to add a space-based node to the EHT to detect and study the photon ring~\cite{2022-Kurczynski-}. Advancements in imaging techniques such as super-resolution and hybrid imaging are expected to improve the resolution to 5 $\mu as$ by the current estimates~\cite{2023-Roelofs-Galaxies}. This, along with the higher-frequency observations (such as 345~GHz), can improve the coverage, along with the improvements in the data analysis techniques can enhance the instrumental resolution in the near future~\cite{2019-VLBI-Galaxies,2020-Johnson-Sci.Adv.,2020-Broderick-apj,2021-Raymond.etal-ApJS,2021-Hadar-Phys.Rev.D}. 

Analyzing the difference between the shadows corresponding to the $'+'$ and $'-'$ modes requires more information regarding their polarization, which {we plan} to analyze in a future work, along with the effect of NMC on the black hole shadow for a variety of accretion flow models. For instance, the black hole shadow with NMC for Novikov-Thorne accretion model~\cite{1973-Novikov-} (with low numerical accuracy) is discussed in Ref.~\cite{sup1-}.

\noindent\emph{Conclusions:} 
EEP is not a basic principle of physics but an empirical fact~\cite{2001-Damour-Talk}. Therefore, testing EEP will assess the validity of GR. NMC of electromagnetic fields violates EEP and {their effects} manifest in the strong-gravity regime. Hence, VLBI observations of black hole image provide an opportunity to test NMC in the strong-gravity regime. We explicitly showed that NMC of the electromagnetic field modifies the black hole image in two ways: First, in linear order in the spin parameter $a$, the horizon casts a \emph{shadow} of radius \emph{greater than} $\sqrt{27/4}~r_H$ on the image of the source for one mode. For the other mode, it is \emph{smaller than} $\sqrt{27/4}~r_H$. Second, the brightness and the position of the lensing ring are affected by the non-minimal coupling. For ${\lambda}<0$ and $'+'$ mode (and ${\lambda}>0$ and $'-'$ mode), the lensing ring is more prominent as compared to other cases. Interestingly, the black hole image observations can provide a weak bound on $\lambda$. The observable photon ring does not form for ${\lambda}<-0.5$  for $'+'$ mode and for ${\lambda} > 0.25$ for the $'-'$ mode. EEP constraint we predict is different from the one reported Gravity collaboration~\cite{2019-Amorim-Phys.Rev.Lett.}. Here our interest is to constrain NMC constant while the authors constrain the violation of EEP via atomic transitions.

For Kerr space-time, the dispersion relation of the NMC photon is more complicated and lacks separability in $r$ and $\zeta$. Hence, in the current work, we considered slowly rotating black holes at linear order in the rotational parameter $a$. Our analysis is valid for $a \leq 0.20$ {($\tilde{a} \leq 0.4 M$)}. In order to compare with ngEHT observation for higher $a$, we need to solve Eq.~\eqref{eq:Disper-Kerr} numerically for higher spins. This is currently in progress. 

EHT has detected high linear polarization fractions $(2–15\%)$ and large rotation measures ($> 10^3$) for Sgr $A^*$ and M87~\cite{2021-ALMA-ApJL,2021-EHT-Polariz-ApJ}. EHT has associated linear polarization due to the {strong magnetic field.} 
Our analysis shows that the non-minimal coupling of the electromagnetic field affects the two polarization modes of the photon distinctly. Thus, our analysis suggests that some percentage of the linear polarization seen in EHT observations can be due to NMC. Note that the synchrotron emission can contribute to the polarized image, which can complicate the proposed measurements~\cite{2021-Akiyama-Astrophys.J.}.
Next, we plan to do a detailed numerical analysis to investigate the polarization effects of NMC and possible constraints from future VLBI projects such as ngEHT and Event Horizon Explorer. 

Since NMC introduces direct interaction between curvature and electromagnetic field, our analysis is independent of the accretion flow. Hence, we have considered radially free-falling optically thin light-emitting accretion flow mode. However, to establish the robustness of the NMC signatures on the black hole image, it is imperative to extend the analysis for other types of accretion models, like Keplerian accretion disk, thin infinite accretion disk, and Ion torus~\cite{2005-Beckwith.Done-MNRAS,2011-Vincent.etal-CQG}. This is currently under investigation.

\noindent\emph{Acknowledgments:} The authors thank M. Johnson, V. Marthi, K. Moriyama, D. Palumbo, and K. Sarkar for insights regarding the current and future VLBI observations. The authors thank A. Choudhury, K. Chandra, S. M. Chandran, A. Kushwaha, K. Rajeev, and S. Xavier for comments. JPJ was funded by IITB-IoE grant. The work is supported by the SERB-Core Research Grant.

\input{references.bbl}
\clearpage
\begin{widetext}
\noindent\section*{{Supplementary material}}
This document contains details and plots related to the calculations presented in the main text. 
\subsection{Photon circular orbit in the Schwarzschild space-time}
\label{app:Schw-Perturb}
The two conditions for the circular orbit of the photon ($\dot{r} = 0$, $\ddot{r}=0$) leads to the following equations:
\begin{align}
      {E}^{2}\, \left[1 - \frac{{b}^2}{{r}^2} \left(1 - \frac{1}{{r}}\right)\,\left(1 -\,\mathcal{C}_{\pm} \right) \right] &= 0, \\
\dfrac{b^2E^2}{2r^4}\left[ (2r-3)(1-\mathcal{C}_{\pm}) + r(r-1)\dfrac{\partial \mathcal{C}_{\pm}}{\partial r} \right] &= 0.
\end{align}
To find the analytical solutions for the radius of the photon circular orbit and the corresponding impact parameter, one can use the perturbative approach. We expand the coordinate ${r}$, and the parameters ${E}$ and ${b}$ in the powers of the coupling constant ${\lambda}$ given by
\begin{equation}
    {r}({s},{\lambda}) = \sum_{i=0}^{n} {r}_{i}({s}){\lambda}^i, \quad  {b}({s},{\lambda}) = \sum_{i=0}^{n} {b}_{i}({s}){\lambda}^i, \quad  
    {E}({s},{\lambda}) = \sum_{i=0}^{n} {E}_{i}({s}){\lambda}^i .
\end{equation}
We can then solve the conditions for the circular orbits order by order to obtain the radius of photon sphere and the corresponding impact parameters as a power series expansion given by
\begin{equation}
    {r}_c^{\pm}({\lambda}) = \sum_{i=0}^{n} {r}^{\pm}_{i}\lambda^i, \quad
    {b}_c^{\pm}({\lambda}) = \sum_{i=0}^{n} {b}^{\pm}_{i}\lambda^i .
\end{equation}
\noindent
For $n=4$, the expansion is given below. For $'+'$ mode,
\begin{eqnarray}
 \label{eq:p_rb_sch}
    && r_c^+(\lambda) = \dfrac{3}{2} + \dfrac{4}{3}\lambda  +  \dfrac{64}{81} \lambda^2 - \lambda^3 \dfrac{2816}{729} - \dfrac{102400}{59049} \lambda^4,  \\ 
    && b_c^+(\lambda) = \dfrac{3\sqrt{3}}{2} + \dfrac{4}{\sqrt{3}}\lambda  -  \dfrac{64}{27\sqrt{3}} \lambda^2 - \dfrac{1280}{729\sqrt{3}}  \lambda^3 + \dfrac{167936}{19683\sqrt{3}}\lambda^4 .
 \nonumber  
\end{eqnarray}
Similarly, for $'-'$ mode
\begin{eqnarray}
     \label{eq:m_rb_sch}
    && r_c^-(\lambda) = \dfrac{3}{2} - \dfrac{4}{3}\lambda  -  \dfrac{64}{81} \lambda^2 - \lambda^3 \dfrac{256}{729} - \dfrac{45056}{59049} \lambda^4 ,\\ 
    && b_c^-(\lambda) = \dfrac{3\sqrt{3}}{2} - \dfrac{4}{\sqrt{3}}\lambda  -  \dfrac{128}{27\sqrt{3}} \lambda^2 - \dfrac{6400}{729\sqrt{3}}  \lambda^3 + \dfrac{389120}{19683\sqrt{3}}\lambda^4 .
    \nonumber
\end{eqnarray}
Deviation of the analytical solution obtained using the perturbative approach from the exact numerical solution is given in \ref{fig:rb_dev}.
For $|\lambda| < 0.2$,  we see that the above perturbative expansion up to $n=4$ is consistent with the exact numerical solution. The deviation is of the order of $\sim 10^{-2}$. For higher values of $\lambda$, perturbative expansion matches with the numerical results for higher values of $n$.
\begin{figure}[!htb]
\includegraphics[scale=0.60]{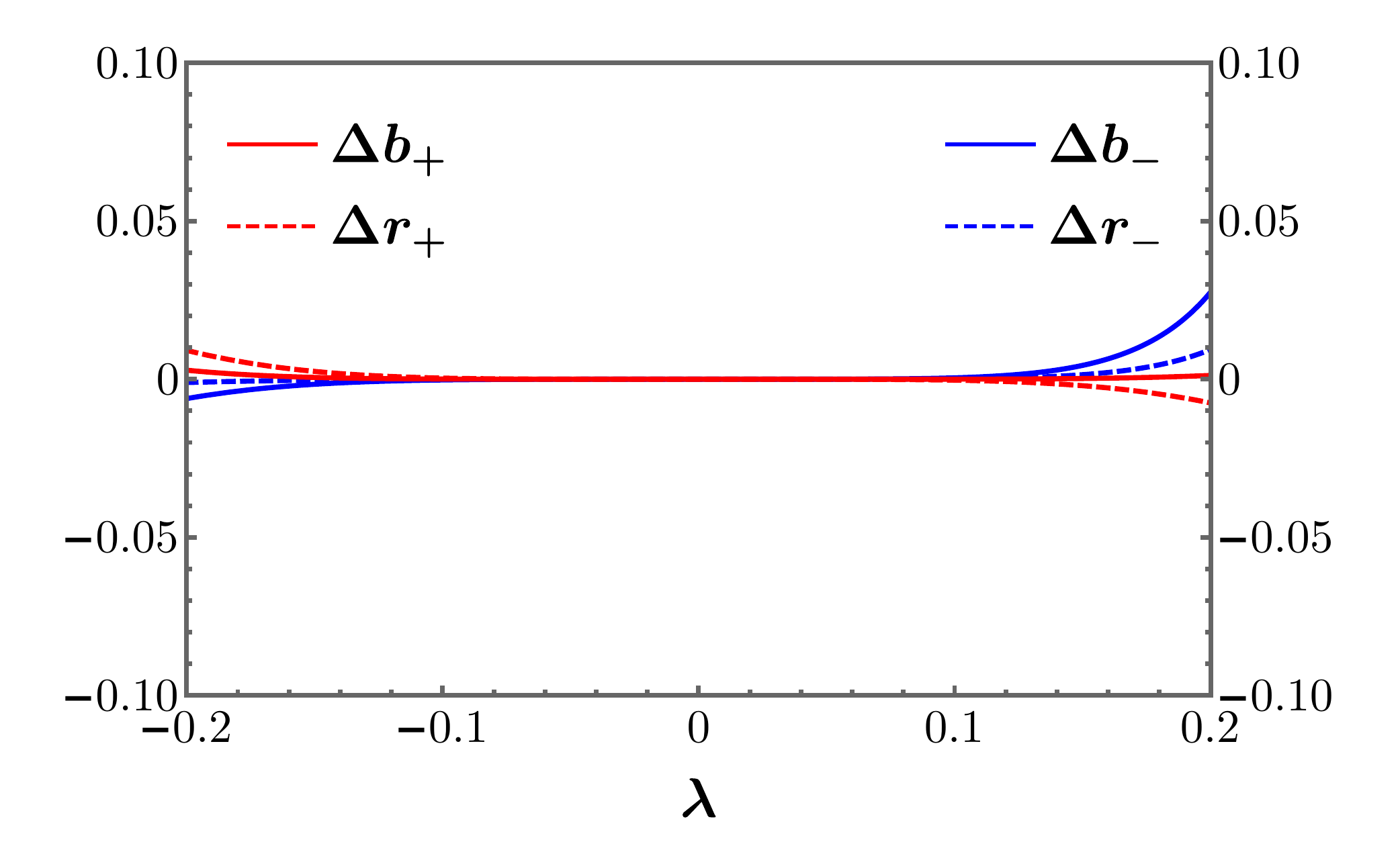}
\caption{Deviation of the perturbed analytical solution from the exact numerical solution for $r_c$ and $b_c$ for $'\pm'$ modes.}
\label{fig:rb_dev}
\end{figure}

\begin{figure}[!htb]
\begin{minipage}[b]{.31\textwidth}
\includegraphics[scale=0.23]{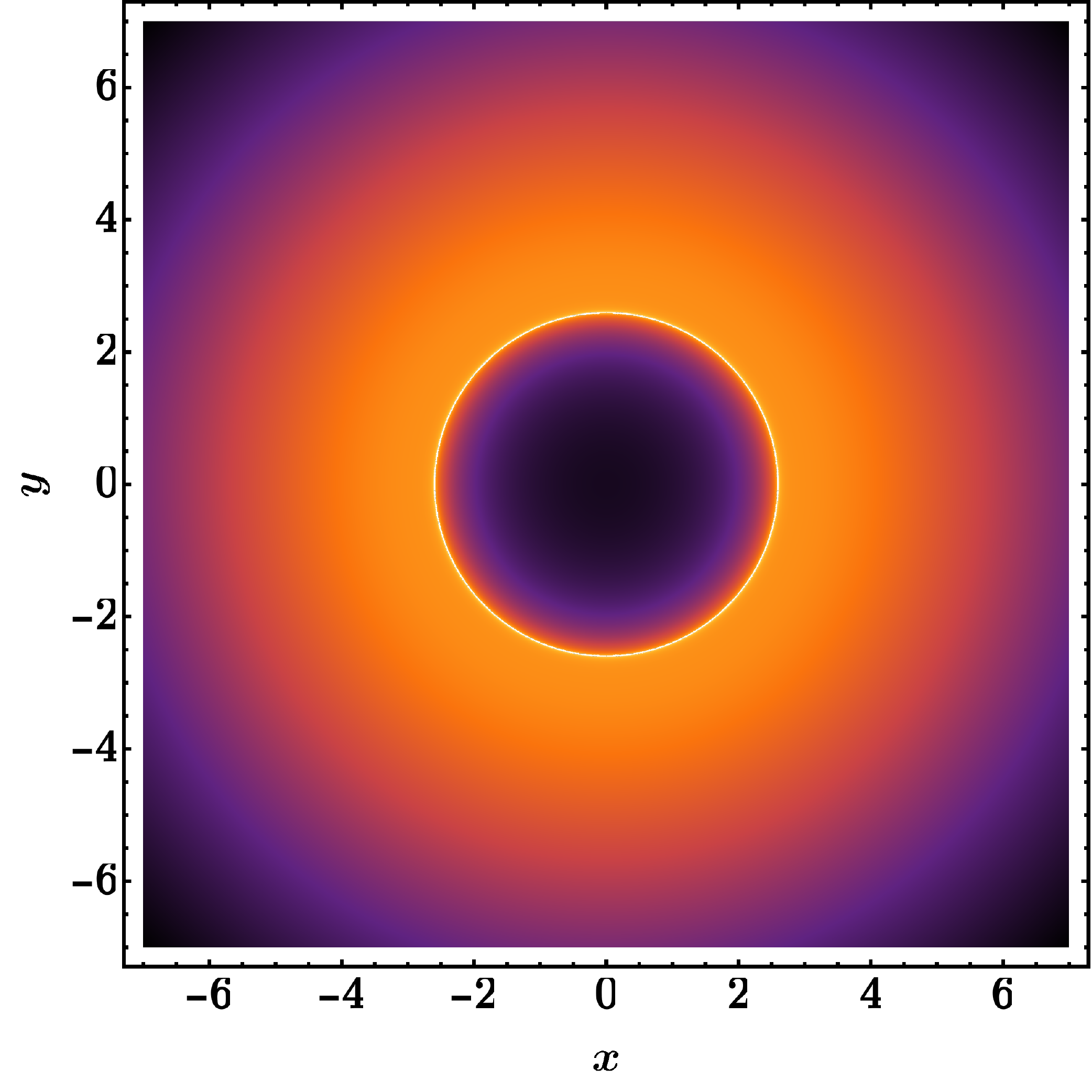}
\end{minipage}\hfill
\begin{minipage}[b]{.31\textwidth}
\includegraphics[scale=0.23]{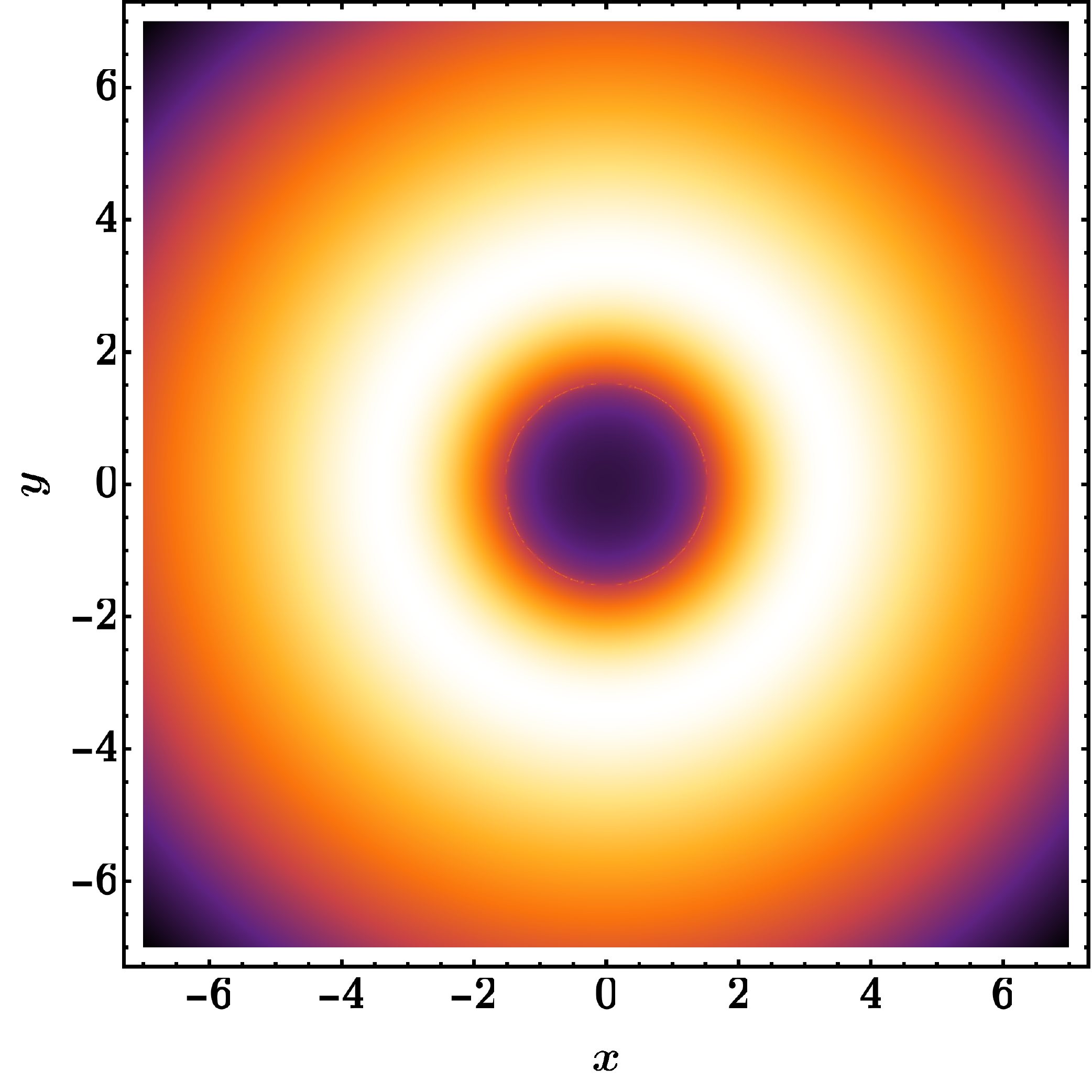}
\end{minipage}\hfill
\begin{minipage}[b]{.31\textwidth}
\includegraphics[scale=0.23]{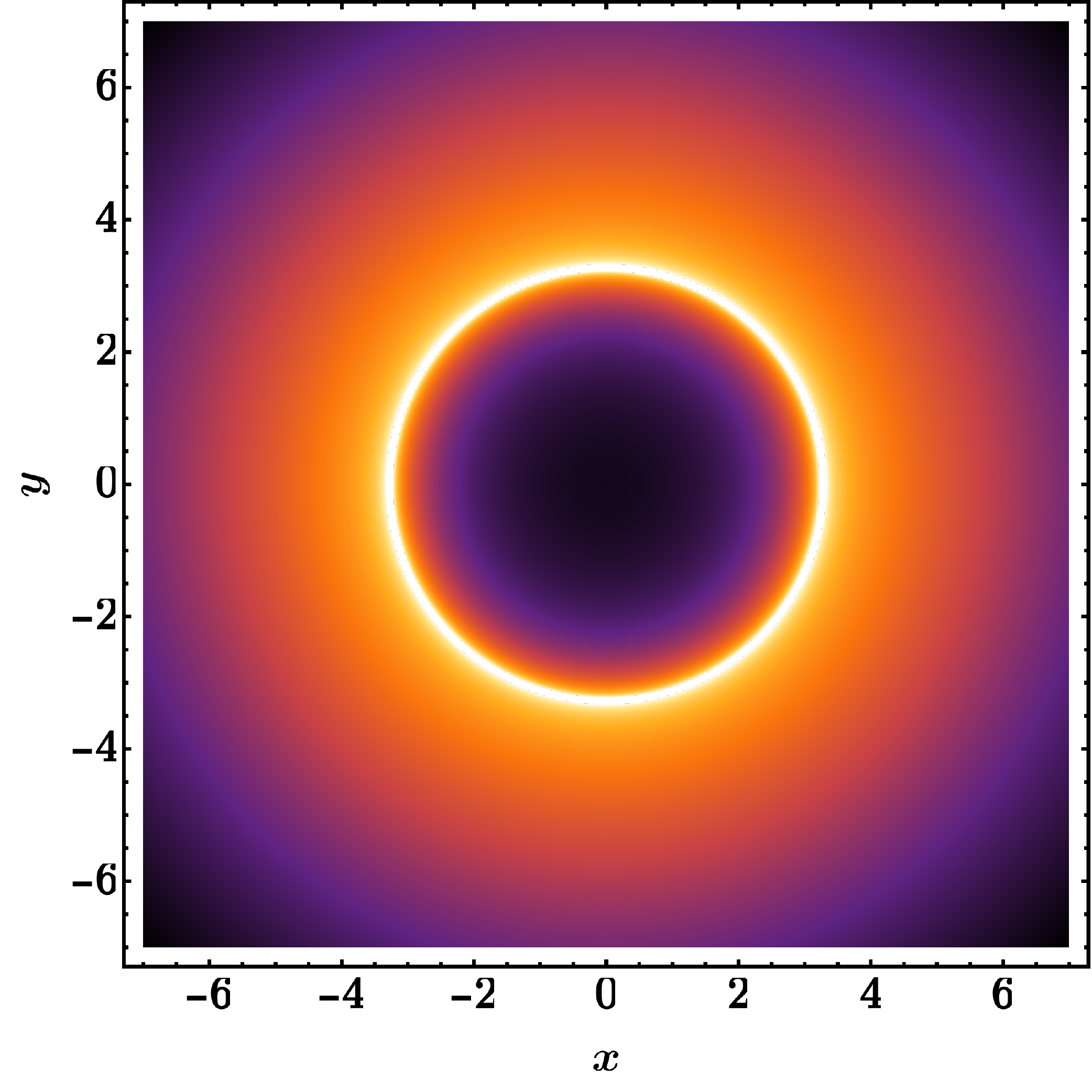}
\end{minipage}
\end{figure}
%
\begin{figure}[!htb]
\begin{minipage}[b]{.30\textwidth}
\includegraphics[scale=0.23]{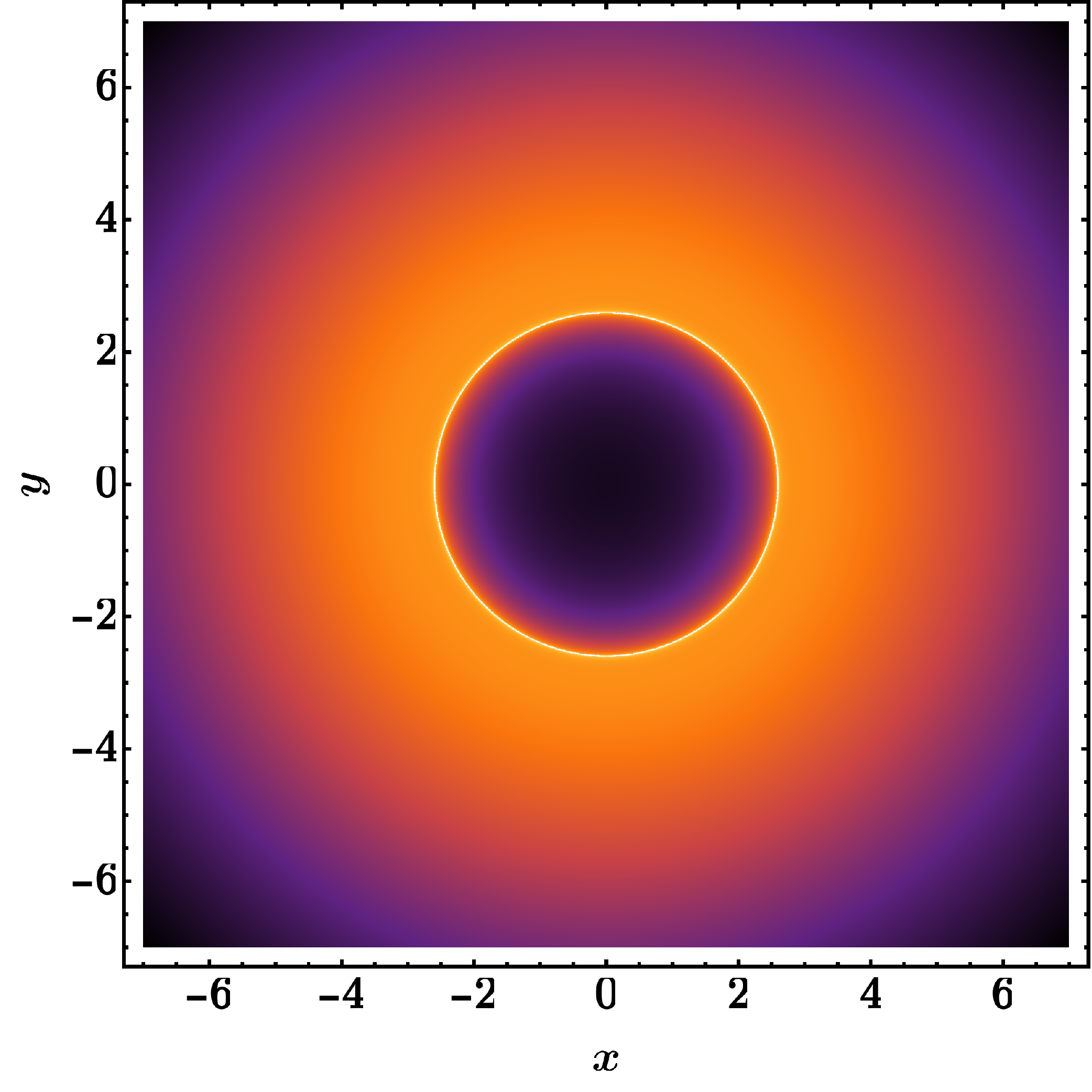}
\end{minipage}\hfill
\begin{minipage}[b]{.30\textwidth}
\includegraphics[scale=0.23]{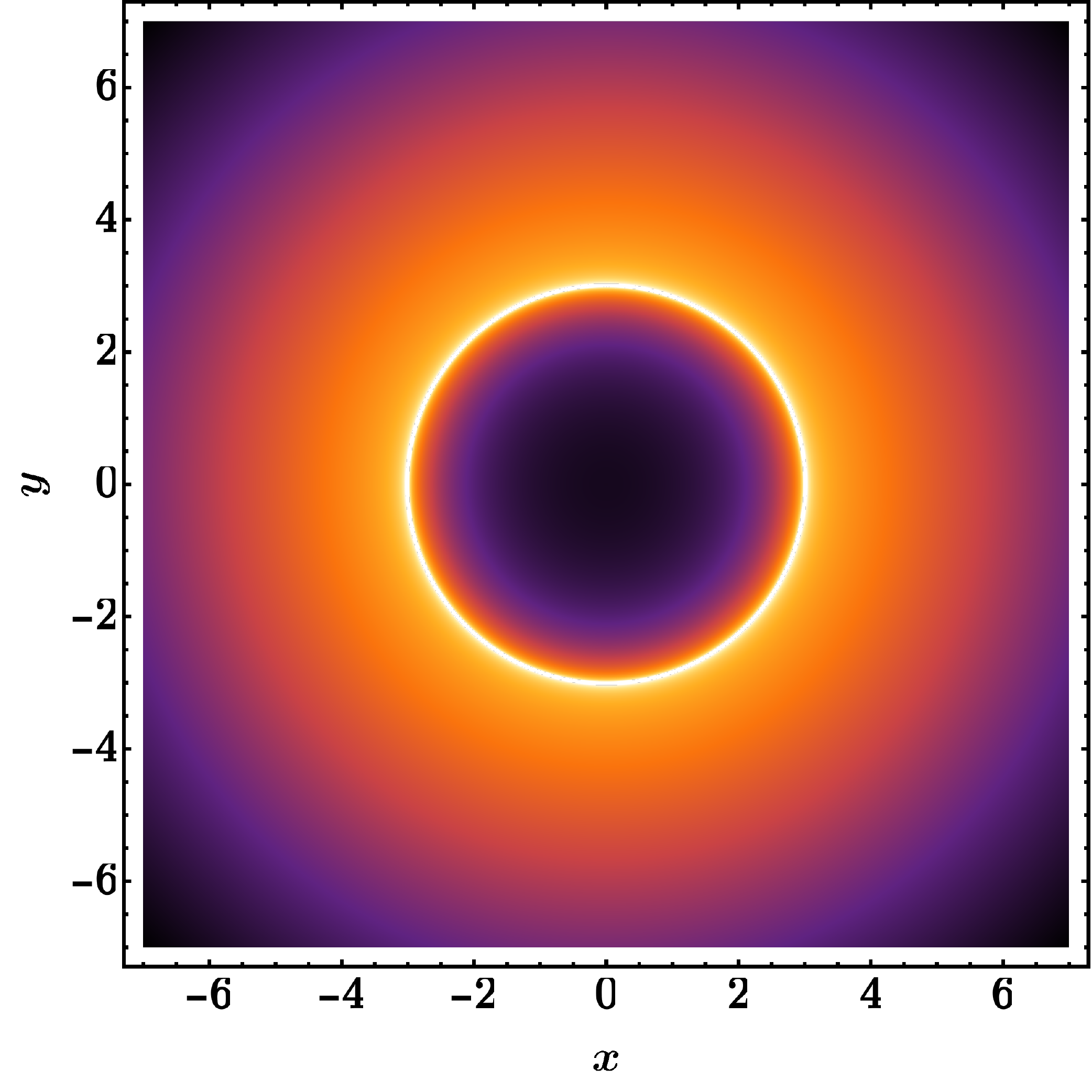}
\end{minipage}\hfill
\begin{minipage}[b]{.30\textwidth}
\includegraphics[scale=0.23]{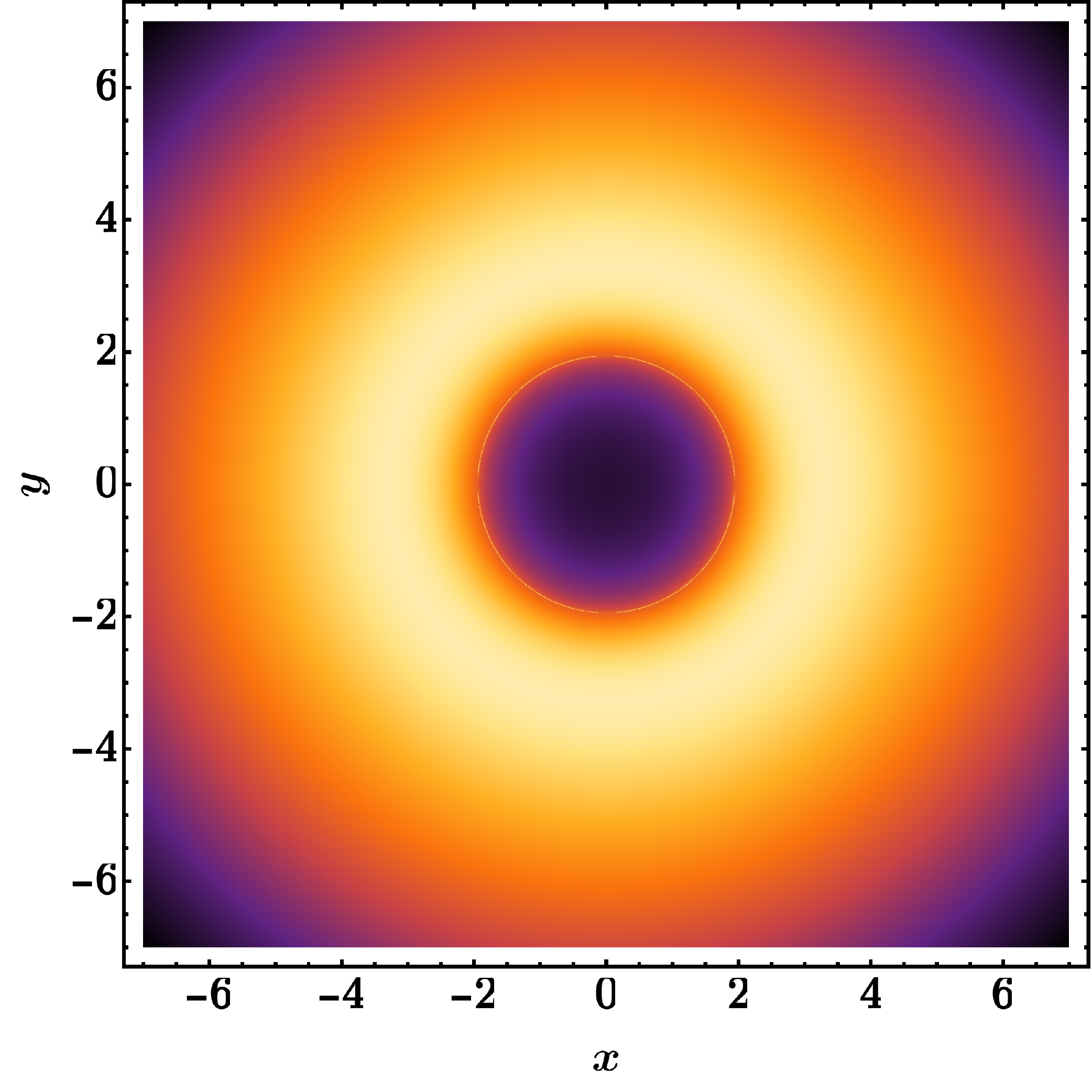}
\end{minipage}
\caption{ $\tilde{\lambda}=-0.4$(top)$, \tilde{\lambda}=0.2$(bottom). Left: $\tilde{\lambda}=0$, Center: `$+$' mode, Right: `$-$' mode.}
\label{fig:BH_image_RI}
\end{figure}

\subsection{Observed image of accretion flow}
For the radially infalling accretion flow described by eq. 10 in the manuscript, the image of the black hole appears as given in \ref{fig:BH_image_RI}. Similarly, for a Novikov-Thorn accretion disk~\cite{1973-Novikov-} with an inclination angle of $60^{o}$, the images for different values of coupling constant are given in \ref{fig:BH_image_NT}. Here the light emitting particles forms a very thin disk and follows near-Keplerian orbits till the inner most stable circular orbit, followed by plunging into the black holes.

Here we see that for the radially infalling accretion flow, the shadow of the black hole is bordered by the photon ring, which is not the case for the Novikov-Thorn accretion disk with the inner radius $r_i = r_{ISCO}$

Note that for the Novikov-Thorne accretion disk in the equatorial plane, we  need to run the ray tracing code with a higher resolution to get the complete image without any gaps. 
Images given here represents the observed shape of the accretion disk. They do not contain any information regarding the observed flux.
\begin{figure}[!htb]
\begin{minipage}[b]{.31\textwidth}
\includegraphics[scale=0.45]{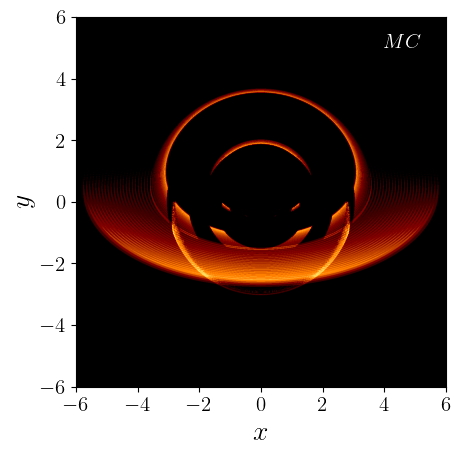}
\end{minipage}\hfill
\begin{minipage}[b]{.31\textwidth}
\includegraphics[scale=0.45]{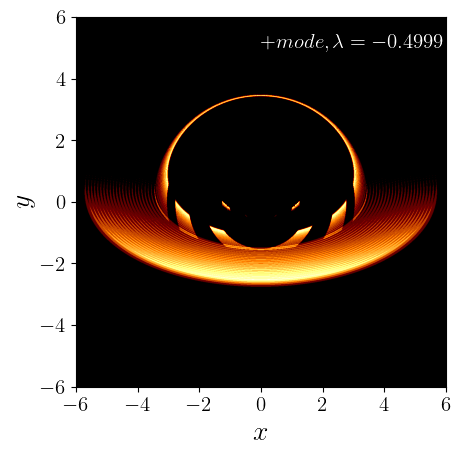}
\end{minipage}\hfill
\begin{minipage}[b]{.31\textwidth}
\includegraphics[scale=0.45
]{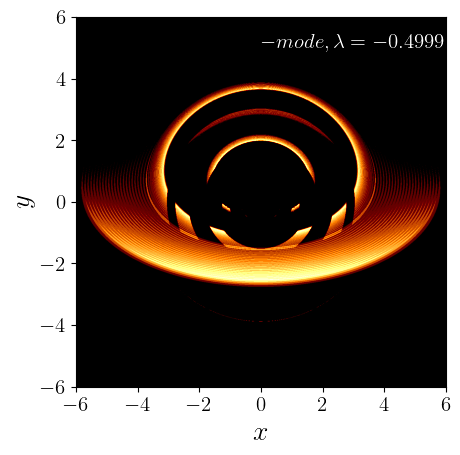}
\end{minipage}
\end{figure}
\begin{figure}[!htb]
\begin{minipage}[b]{.31\textwidth}
\includegraphics[scale=0.45]{mc_image.png}
\end{minipage}\hfill
\begin{minipage}[b]{.31\textwidth}
\includegraphics[scale=0.45]{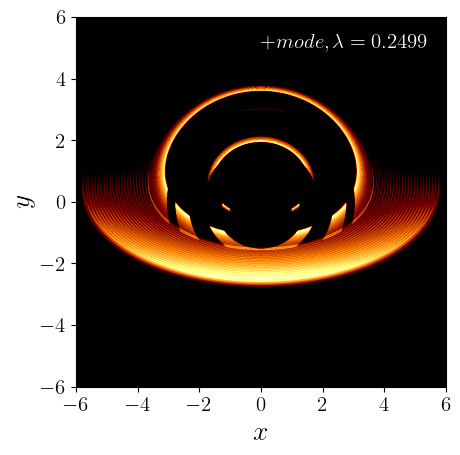}
\end{minipage}\hfill
\begin{minipage}[b]{.31\textwidth}
\includegraphics[scale=0.45]{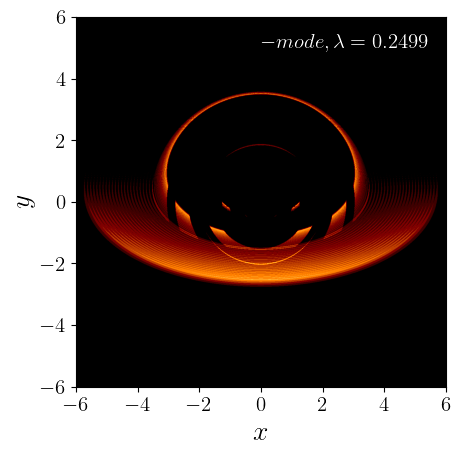}
\end{minipage}
\caption{ $\tilde{\lambda}=-0.4999$(top)$, \tilde{\lambda}=0.2499$(bottom). Left: $\tilde{\lambda}=0$, Center: `$+$' mode, Right: `$-$' mode.}
\label{fig:BH_image_NT}
\end{figure}
\subsection{Photon circular orbit in the equatorial plane of a Kerr black hole}
\label{app:Kerr-Perturb}

%

Radius of photon circular orbit and the corresponding impact parameter ${b}$ can be obtained by imposing the independent conditions $\dot{{r}} = 0$ and $\ddot{{r}}=0$. These equations can be solved numerically to obtain the radius of the circular photon orbit and corresponding impact parameter. However, as in the case of Schwarzschild black hole, here we use the analytical perturbative approach to obtain the radius of the circular photon orbit and the corresponding impact parameter as the function of the coupling constant ${\lambda}$ and the spin ${a}$ as given below. Here each coefficient of ${\lambda}$ is expanded in terms of ${a}$. Depending on the value of ${\lambda}$ and ${a}$ in consideration, one can use the expansion up to an appropriate order $n$.
\begin{eqnarray}
    {r}_c^{\pm}({a},{\lambda}) = \sum_{i,j=0}^{n} {r}^{\pm}_{ij}{a}^i \lambda^j, \quad
    {b}_c^{\pm}({a},{\lambda}) = \sum_{i,j=0}^{n} {b}^{\pm}_{ij}{a}^i \lambda^j .
\end{eqnarray}
The coefficients ${r}^{\pm}_{ij}$ and ${b}^{\pm}_{ij}$ for $n=3$ are the  given by Eqs.(\ref{eq:p_r_kerr},\ref{eq:p_b_kerr},\ref{eq:m_r_kerr},\ref{eq:m_b_kerr}) .

Radius of the circular orbit of the non-minimally coupled photon in the equatorial plane of a Kerr black hole and the corresponding impact parameter can be expanded as
\begin{eqnarray}
    {r}_c^{\pm}({a},{\lambda}) = \sum_{i,j=0}^{4} {r}^{\pm}_{ij}{\lambda}^i {a}^j, \quad
    {b}_c^{\pm}({a},{\lambda}) = \sum_{i,j=0}^{4} {b}^{\pm}_{ij}{\lambda}^i {a}^j .
\end{eqnarray}
For $'+'$ mode, the coefficients are given by
\begin{eqnarray}\nonumber
   && r^+_{00} = \dfrac{3}{2}, \quad r^+_{01}=-\dfrac{2}{\sqrt{3}}, \quad r^+_{02} = -\dfrac{4}{9}, \quad r^+_{03} = -\dfrac{20}{27\sqrt{3}} ,
\\ \nonumber
   && r^+_{10} = \dfrac{4}{3}, \quad r^+_{11} = -\dfrac{32}{9\sqrt{3}}, \quad r^+_{12} = -\dfrac{16}{9}, \quad r^+_{13} = -\dfrac{256}{81\sqrt{3}} ,
   \\  \label{eq:p_r_kerr} 
   && r^+_{20} = \dfrac{64}{81}, \quad r^+_{21} = \dfrac{1280}{81\sqrt{3}}, \quad r^+_{22}=-\dfrac{8704}{2187}, \quad r^+_{23} = -\dfrac{248576}{6561\sqrt{3}}, 
   \\ 
   &&  { r^+_{30} = -\dfrac{2816}{729}, \quad r^+_{31} = \dfrac{57344}{6561\sqrt{3}}, \quad r^+_{32}=-\dfrac{7737344}{59049}, \quad r^+_{33} = -\dfrac{16310272}{59049\sqrt{3}}}.
  \nonumber
\end{eqnarray}
\begin{eqnarray}
\label{eq:p_b_kerr}
   && b^+_{00} = \dfrac{3\sqrt{3}}{2}, \quad b^+_{01}=-2, \quad b^+_{02} = -\dfrac{1}{\sqrt{3}}, \quad b^+_{03} = -\dfrac{16}{27} ,
\\ \nonumber
   && b^+_{10} = \dfrac{4}{\sqrt{3}}, \quad b^+_{11} = -\dfrac{8}{9\sqrt{3}}, \quad b^+_{12} = -\dfrac{256}{243}, \quad b^+_{13} = -\dfrac{88}{27\sqrt{3}} ,
   \\ \nonumber  
   && b^+_{20} = -\dfrac{64}{27\sqrt{3}}, \quad b^+_{21} = \dfrac{512}{81}, \quad b^+_{22}=\dfrac{3520}{243\sqrt{3}}, \quad b^+_{23} = \dfrac{5120}{729\sqrt{3}} ,
   \\ 
   &&  {b^+_{30} = -\dfrac{1280}{729\sqrt{3}}, \quad b^+_{31} = -\dfrac{40960}{2187}, \quad b^+_{32}=\dfrac{28672}{729\sqrt{3}}, \quad b^+_{33} = \dfrac{19382272}{177147}} .
   \nonumber
\end{eqnarray}

For $'-'$ mode
\begin{eqnarray}\nonumber
   && r^-_{00} = \dfrac{3}{2}, \quad r^-_{01}=-\dfrac{2}{\sqrt{3}}, \quad r^-_{02} = -\dfrac{4}{9}, \quad r^-_{03} = -\dfrac{20}{27\sqrt{3}} ,
\\ \nonumber
   && r^-_{10} = -\dfrac{4}{3}, \quad  r^-_{11} = \dfrac{32}{9\sqrt{3}}, \quad r^-_{12} = \dfrac{16}{9}, \quad r^-_{13} = \dfrac{256}{81\sqrt{3}} ,
   \\ \label{eq:m_r_kerr}
   && r^-_{20} = -\dfrac{64}{81}, \quad r^-_{21} = \dfrac{1024}{81\sqrt{3}}, \quad r^-_{22}=-\dfrac{11264}{2187}, \quad r^-_{23} = -\dfrac{239872}{6561\sqrt{3}} ,
   \\ 
   && {r^-_{30} = -\dfrac{256}{729}, \quad r^-_{31} = \dfrac{139264}{6561\sqrt{3}}, \quad r^-_{32}=-\dfrac{4911104}{59049}, \quad r^-_{33} = -\dfrac{9699328}{59049\sqrt{3}}} .
   \nonumber
\end{eqnarray}
\begin{eqnarray}
 \label{eq:m_b_kerr}
   && b^-_{00} = \dfrac{3\sqrt{3}}{2}, \quad b^-_{01}=-2, \quad  b^-_{02} = -\dfrac{1}{\sqrt{3}}, \quad b^-_{03} = -\dfrac{16}{27} ,
\\ \nonumber
   && b^-_{10} = -\dfrac{4}{\sqrt{3}}, \quad b^-_{11} =0 , \quad b^-_{12} = \dfrac{8}{9\sqrt{3}}, \quad b^-_{13} = \dfrac{256}{243} ,
   \\ \nonumber
   && b^-_{20} = -\dfrac{128}{27\sqrt{3}}, \quad b^-_{21} = \dfrac{256}{81}, \quad b^-_{22}=\dfrac{1088}{243\sqrt{3}}, \quad b^-_{23} = -\dfrac{2048}{729} ,
   \\ 
   &&  {b^-_{30} = -\dfrac{6400}{729\sqrt{3}}, \quad b^-_{31} = \dfrac{28672}{2187}, \quad  b^-_{32}=-\dfrac{7168}{243\sqrt{3}}, \quad b^-_{33} = \dfrac{13828096}{177147}} .
   \nonumber
\end{eqnarray}
\subsection{Observed shadow of the Kerr black hole}
\ref{fig:kerr_shadow} shows the outline of the observed shadow of the Kerr black hole for non-minimally coupled modes and minimally coupled mode for different values of spin ${a}$.
\begin{figure}[!htb]
\includegraphics[scale=0.8]{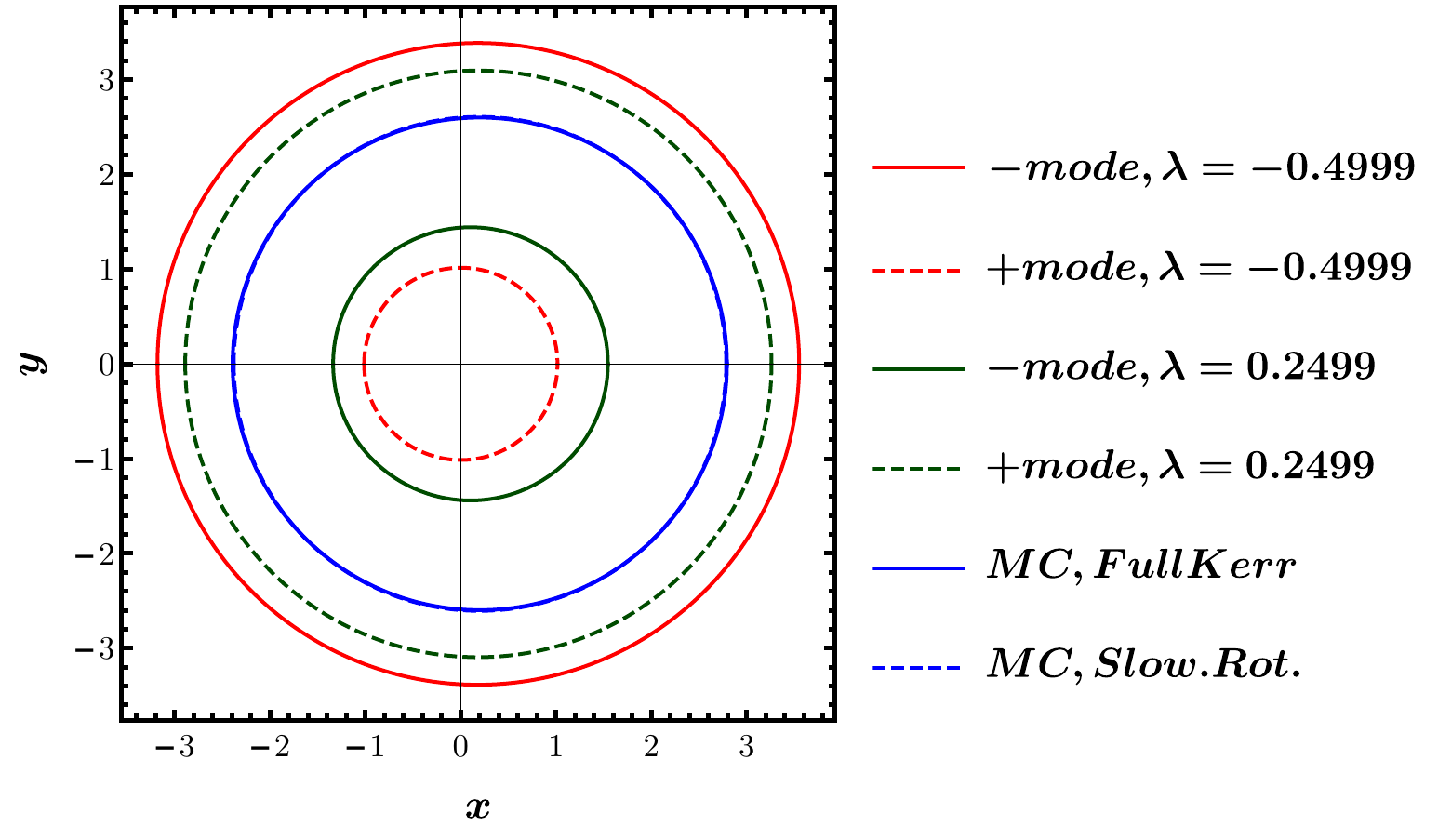}
\includegraphics[scale=.801]{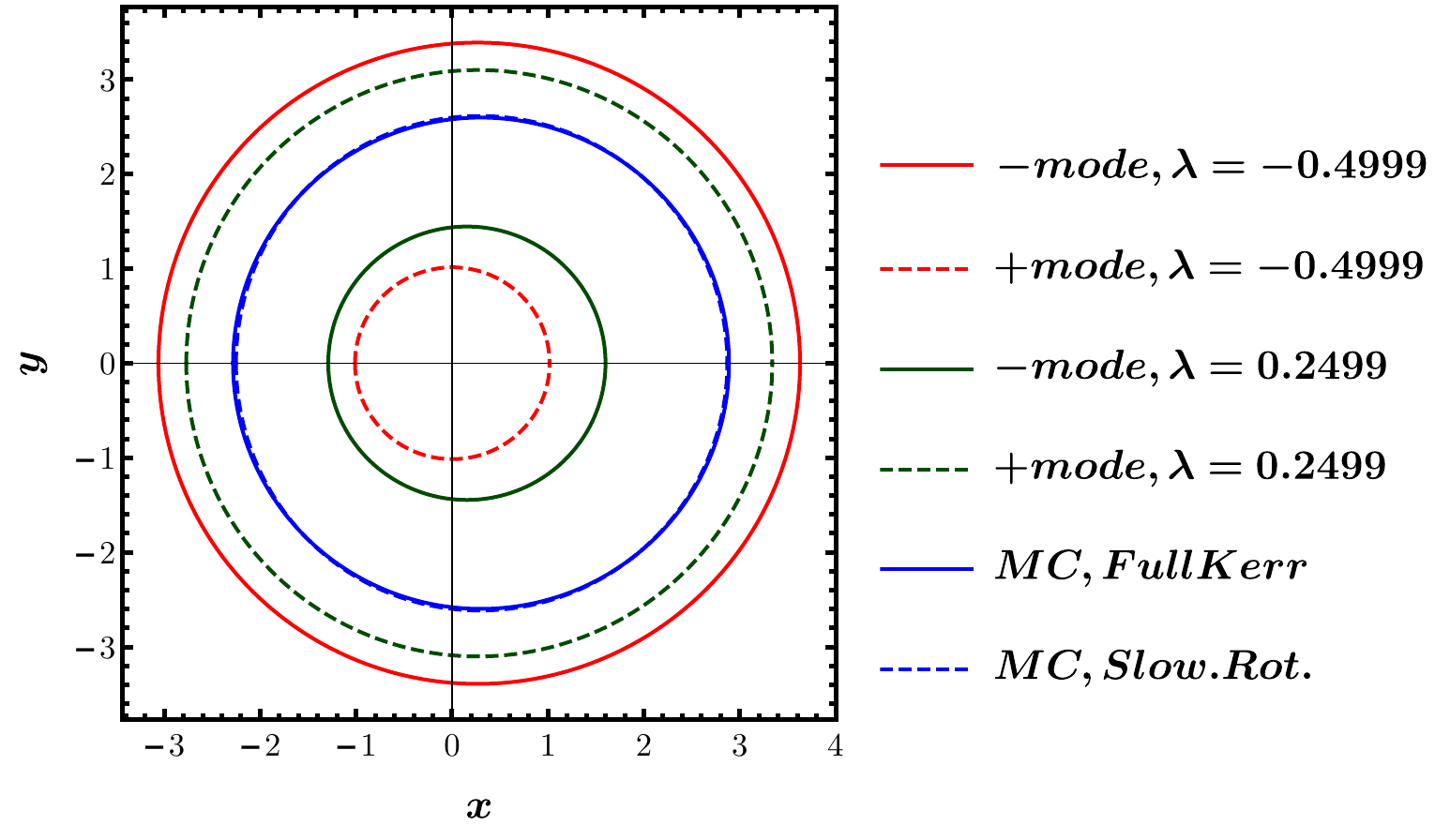}
\caption{Outline of the shadow of a slowly rotating black hole observed by a distant LNRO for different $\pm$ photon modes for near limiting values of ${\lambda}$ along with the minimally coupled case for ${a}=0.1$ (top) and ${a}=0.15$ (bottom).}
\label{fig:kerr_shadow}
\end{figure}
\end{widetext}
\end{document}

%% file: references.bbl
%